\documentclass[12pt]{article}

\usepackage[letterpaper, margin=0.5in]{geometry}
\usepackage{float}
\usepackage{graphicx}
\usepackage{lineno}
\usepackage{url}
\usepackage{amsfonts}
\usepackage{amsmath}  
\usepackage{array}

\usepackage{ulem}
\usepackage{authblk}
\usepackage[utf8]{inputenc}

\title{Beam Profiler Network (BPNet)- A Deep Learning Approach to Mode Demultiplexing of Laguerre-Gaussian Optical Beams}

\author[1, $\ddagger$]{Amit Bekerman}
\author[1, $\ddagger$]{Sahar Froim}
\author[1, $\ddagger$]{Barak Hadad}
\author[1,*]{Alon Bahabad}

\affil[1]{Department of Physical Electronics, School of Electrical Engineering, Fleischman Faculty of Engineering, Tel-Aviv University, Tel-Aviv 69978, Israel}
\affil[$\ddagger$]{Equal contribution}
\affil[*]{Corresponding author: alonb@eng.tau.ac.il}

\begin{document}
	
	\maketitle
	
	\begin{abstract}
		The transverse field profile of light is being recognized as a resource for classical and quantum communications for which reliable methods of sorting or demultiplexing spatial optical modes are required. 
		Here, we demonstrate, experimentally, state-of-the-art mode demultiplexing of Laguerre-Gaussian beams according to both their orbital angular momentum and radial topological numbers using a flow of two concatenated deep neural networks. The first network serves as a transfer function from experimentally-generated to ideal numerically-generated data, while using a unique "Histogram Weighted Loss" function that solves the problem of images with limited significant information. The second network acts as a spatial-modes classifier. Our method uses only the intensity profile of modes or their superposition, making the phase information redundant.
	\end{abstract}

	\section{Introduction}
	The possibility of employing the spatial degree of photons for communications is gaining interest in recent years due to its unbounded dimensionality \cite{wang2012terabit,nagali2009quantum,bozinovic2013terabit,krenn2014generation}. A natural basis to span the transverse profile of photons is comprised of Laguerre-Gaussian (LG) modes which are characterized with two topological numbers: $l\in \mathbb{Z}$, the orbital index,  describing the orbital-angular-momentum (OAM) in units of $\hbar$ per photon in the beam and $p\in \mathbb Z_+$, which is the radial index or radial quantum number. Essential for utilizing LG modes is the ability to perform mode-sorting, or demultiplexing on the incoming physical data flow. There are, essentially, two approaches to demultiplexing. The first approach uses (usually complicated) optical setups in which the $l$ and $p$ degrees of freedom are coupled to other degrees of freedom such as angle of propagation and polarization. Most such methods address either the OAM \cite{lightman2017miniature, doster2017machine, lohani2018use} or the radial index \cite{gu2018gouy, zhou2017sorting} degrees of freedom, while a recent measuring method handles both degrees \cite{bouchard2018measuring}. The second approach, which emerged recently, suggests using just a camera to detect the intensity of the incoming light beam and to utilize a deep neural network (DNN) to classify the beam. To date, demonstrated DNN-based demultiplexers addressed solely the OAM degree of light.
	
	Here, we present, experimentally, a DNN-based mode sorting of both topological numbers of LG modes able to classify both the OAM and radial index. Our solution uses two concatenated DNNs. One network is used for mode classification and it is trained on numerically generated "ideal"  images of LG modes and two-modes superpositions. The other network is a calibration network which converts experimentally detected images, that suffer from optical aberrations and noise, to ideal numerical images, which are then fed to the classifying network. 
	
	\section{Methods}
	
	\subsection{Data generation}
	
	Laguerre-Gauss modes are solutions of the paraxial wave equation in cylindrical coordinates. They are given with  \cite{gu2018gouy}:

	\begin{multline}
	LG_{l,p} (r, \phi, z) = \sqrt{\frac{2 p!}{\pi(p + |l|)!}} \frac{1}{w_z} \bigg(\frac{\sqrt{2}r}{w_z}\bigg)^{|l|} L_p^{|l|} \bigg(\frac{2r^2}{w_z^2}\bigg) \\ exp \bigg( -\frac{r^2}{w_z^2} + i (\frac{kr^2}{2 R_z} + l\phi - (2p + |l| + 1) \varphi_g)\bigg),
	\label{eq:lauggere_gauss_mode}
	\end{multline}

 	where $l$ and $p$ are the orbital and radial indices respectively, $L_p^{|l|}$ are the Laguerre polynomials, $w_z = w_0 \sqrt{1 + (z/z_R)^2}$ is the beam waist with $w_0$ being the waist at $z=0$, $z_R = (\pi w_0^2)/ \lambda$ is the Rayleigh range , $R_z = z(1 + (z_R/z)^2)$ is the radius of curvature, $\lambda$ is the wavelength, $k=2\pi/\lambda$ is the wave number and $\varphi_g = arctan(z/z_R)$ is the Gouy phase.  
	
	Our work uses both numerically and experimentally generated LG modes and their superpositions. 
	Experimentally generated data was acquired in a setup consisting of a 532nm CW laser (Quantum Ventus 532 Solo Laser) which is expanded and collimated before reflecting off a phase-only Spatial Light Modulator (Holoeye Pluto SLM). The phase masks loaded onto the SLM were encoded by extracting the phase of our numerically generated suprimposed modes and then adding a blazed grating to it. The resulting image in the first order of diffraction of the grating is Fourier transformed using a 50cm lens and imaged by a camera (DataRay WinCamD-LCM4). The experimentally generated data is different than the numerically generated data of ideal modes and their superposition due to  inherent aberrations and noise in the optical system.

	Different datasets, each with a different number of superimposed modes were created. We mark the datasets with $DB_N^{type}$ where $N=1,2,3$ is the number of superimposed modes and $type \in \{num,exp\}$ stands for numerically or experimentally generated dataset. Each member of the datasets was realized according to  $\frac{1}{\sqrt{N}} \sum_{n=1}^{N}LG_{l_n,p_n}$ where $l_n$ and $p_n$ are the orbital and radial indices, that are being used in a particular superposition. All images are set to be of size 224x224 pixels with 256 values per pixel in the range of [-1,1] while for the $exp$ datasets the average pixel value (between all images) was set to 0 and the variance to 1 (we note that  experimentally acquired pictures are first obtained at a size of 500x500 with pixel values in the range of [0,255]). 
	Initially, the datasets $DB_N^{num}, N=1,2,3$, contain each $36^N$  members, while $DB_2^{num}$  contains $DB_1^{num}$ and $DB_3^{num}$ contains both $DB_2^{num}$ and $DB_1^{num}$. For training a DNN, very large datasets are required. For this reason the datasets are "augmented" with new members generated from the old ones. Specifically for $DB_1^{num}$ and $DB_2^{num}$ we augment the basic training dataset of images by adding 70 image variation per image in the basic set through beam rotations (Rotation angles were uniformly distributed over $[0,2\pi]$ with 1 deg$\simeq 17$mrad resolution), beam shifts (uniformly distributed over [0,16] pixel shifts in both x and y coordinates) and addition of Gaussian noise with mean $\mu = 0$ and variance $\sigma^2 = 0.2$. This amounts to overall $\sim2,500$ and $\sim90,000$ samples for $DB_1^{num}$ and $DB_2^{num}$, respectively. In $DB_3$ the number of augmentations per unique combination was $2$ leading a total dataset size of $\sim 90,000$. Similarly, $DB_1^{exp}$ and $DB_2^{exp}$ were generated and numerically augmented 100 fold by applying similar rotations and random noise (with no beam shifts).

	\subsection{Network Architecture}
	Our solution to the Beam Profiler Network (BPNet) consists of two concatenated networks, trained separately: a calibration network and a classifier network. Both networks were created using keras API \cite{chollet2015keras}.
	
	The calibration network, based on U-Net's architecture \cite{ronneberger2015u}, converts LG beam images taken in the lab (both single-mode and superpositions) into ideal images of the same LG beams without changing the overall size of the image. The output of the network for each image, which is also the label during training, is an image of the same LG mode (or superposition of LG modes) albeit an ideal one, created using a simulation.
	
	Since large parts of each image in the dataset are dark, a simple Mean-Square-Error (MSE) or Mean-Absolute-Error (MAE) loss functions cannot allow the calibration network to converge properly as with such loss functions the network converges to a poor local minimum, predicting only dark images. To solve this convergence issue, we introduce here a new type of loss function, called "Histogram Weighted Loss" (HWL). This loss  gives higher significance to pixel values that are less common in the image, since they are the ones carrying important information in sparse images. To implement this cost function we first calculate the histogram of each image and modify the calculation of the regular MAE loss during training, by multiplying the loss of each pixel by 1 minus the pixel probability in the image (which is determined by the histogram). In this case, a wrong prediction of a less common pixel will have a higher cost. The Histogram Weighted Loss is given by the following equation:
	
	\begin{equation}
	HWL = \frac{1}{N \times M}\sum_{j=1}^{M}\sum_{i=1}^{N}(1-prob_{i,j})^\gamma |y_{i,j} - y_{i,j}^p|
	\label{eq:hist_loss}
	\end{equation}
	
	 Where $N$ is the number of pixels in an image, $M$ is the number of images in a given batch, $y_{i,j}$ is the value of pixel $i$ in the $j$th (target) image in the batch, $y_{i,j}^p$ is the prediction for the value of the same pixel generated by the network, $prob_{i,j}$ is the probability for said pixel value in the target image (extracted from the histogram) and $\gamma=4$ is a hyper-parameter which we found by trial and error to produce the best results. Histogram Weighted Loss is somewhat related to the concept of Focal Loss \cite{lin2017focal} in which whole images are given different significance for a specific loss calculation.

	The classifier network classifies images fed by the calibration network according to the values of the index numbers of the modes comprising the images.  The classifier network is trained using simulated images of LG beams (both single-mode and superpositions). This network is based on Mobilenet V2's architecture \cite{sandler2018mobilenetv2}, where the last fully-connected layer outputs 36 labels ((p,l)=(0,0),(1,0),...(5,5)). We refer to the set of these 36 labels as the "modes vector". Each label output is set in the range [0,1] indicating the  probability for successful detection of a specific mode. In the results section below, we decide on a mode being detected if the appropriate value in the modes vector is higher than 0.5. 
	The input to this network is a numerically calculated image of an LG beam, or a superposition of such beams, and its output is a labeling of the different modes it contains.  
	
	\section{Results}
	The classifier network was first tested in a stand-alone configuration, by being fed directly by numerically generated and augmented datasets $DB_N^{num}, N=1,2,3$.  We split our three datasets into training (85\%) and validation (15\%) sets. The validation scored a perfect 100\% success rate in all three cases. This shows that the network architecture we used was adapted effectively enough so as to learn even small datasets ($DB_1$) and datasets without a lot of repetition in the samples ($DB_3$). It is notable that when noisy but otherwise undistorted images are supplied to the classifier network, it exhibits very high performance. However, when testing the classifier network on experimental data that was measured in the lab, the inherent aberrations (in any optical system) degrades the performance considerably (see e.g. Ref.\cite{lohani2018use}). At this point we can choose between two strategies. One option is training the classifier network on experimental data (a strategy that was adapted for example in Ref.\cite{doster2017machine}) which has two problems - the performance would rely on the amount of distortions (aberrations) in the optical system and the solution is applicable to a specific optical setup. The second strategy, that we chose, is using a calibration network which let us use a high performance classifier network that could work in principal with any setup-specific calibration network. The performance of the whole system thus becomes dependent mostly on the quality of the calibration network, whose performance depends in turn on the quality of the optical system.

	The next stage was training the calibration network and then testing the whole BPNet (calibration+classifier). For this purpose we divided $DB_1^{exp}$ and  $DB_2^{exp}$ to a training set (72\%), a validation set (18\%) and to a test set  (10\%). The whole network was tested in several different training configurations as described in Table.\ref{table: results}. The main conclusions from these results is that we were able to get state-of-the-art real-world single-mode detection and two-mode superposition demultiplexing. It is noteworthy that single-mode training yield a perfect performance although the dataset for training was relatively small.  The relatively low performance for demultiplexing two-state superpositions when the classifier network was trained with three-term superpositions is attributed to the small level of augmentation in $DB_3^{num}$.

\begin{figure}[htbp]
		\centering
		\includegraphics[width=1\linewidth,trim={3cm 8cm 4cm 7.3cm},clip]{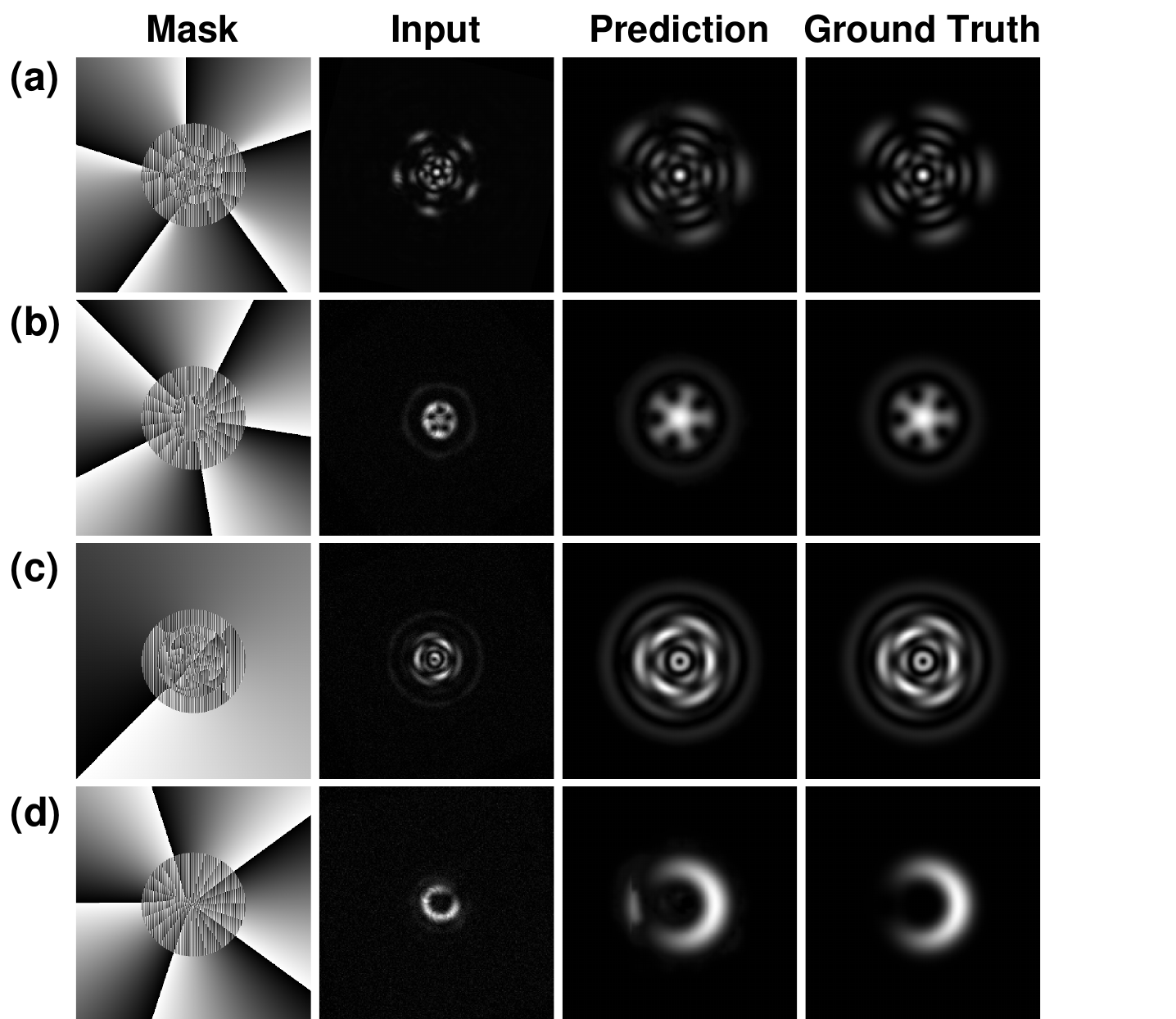}
		\caption{\textbf{Successful classification of two-modes superpositions:} Some examples of successful calibration and classification of modes using the whole BPNet flow. The "Mask" column shows phase masks that where loaded onto the SLM.  The "Input" column shows images captured by the camera. The "Prediction" column shows the output of the calibration network while the "Ground Truth" column shows the projected output for a perfect calibration. The superpositions $((p_1,l_1),(p_2,l_2))$ demonstrated are: \textbf{(a)} ((4,0),(2,5)) \textbf{(b)} ((1,5),(0,0)) \textbf{(c)} ((5,1),(0,4)) \textbf{(d)} ((0,4),(0,5)).
		}
		\label{fig:successful_SP2}
	\end{figure}

	A few examples of successful demultiplexing-detection by the BPNet for two-state superpositions are shown in Fig.\ref{fig:successful_SP2}. In this figure we show the phase masks that were applied to the SLM in the experimental setup, the input to the network which is the image captured by the camera, the prediction of the calibration network which is fed to the classification network, as well as the ground truth for that network. An interesting case is shown in Fig.\ref{fig:successful_SP2}(d) in which the calibration network added an artifact to the prediction but still the resulting classification was correct. This shows that the classiffier network has some degree of robustness to artifacts introduced by the calibration network.

\begin{figure}[htbp]
	\centering
	\includegraphics[width=1\linewidth,trim={3cm 8cm 4cm 7.3cm},clip]{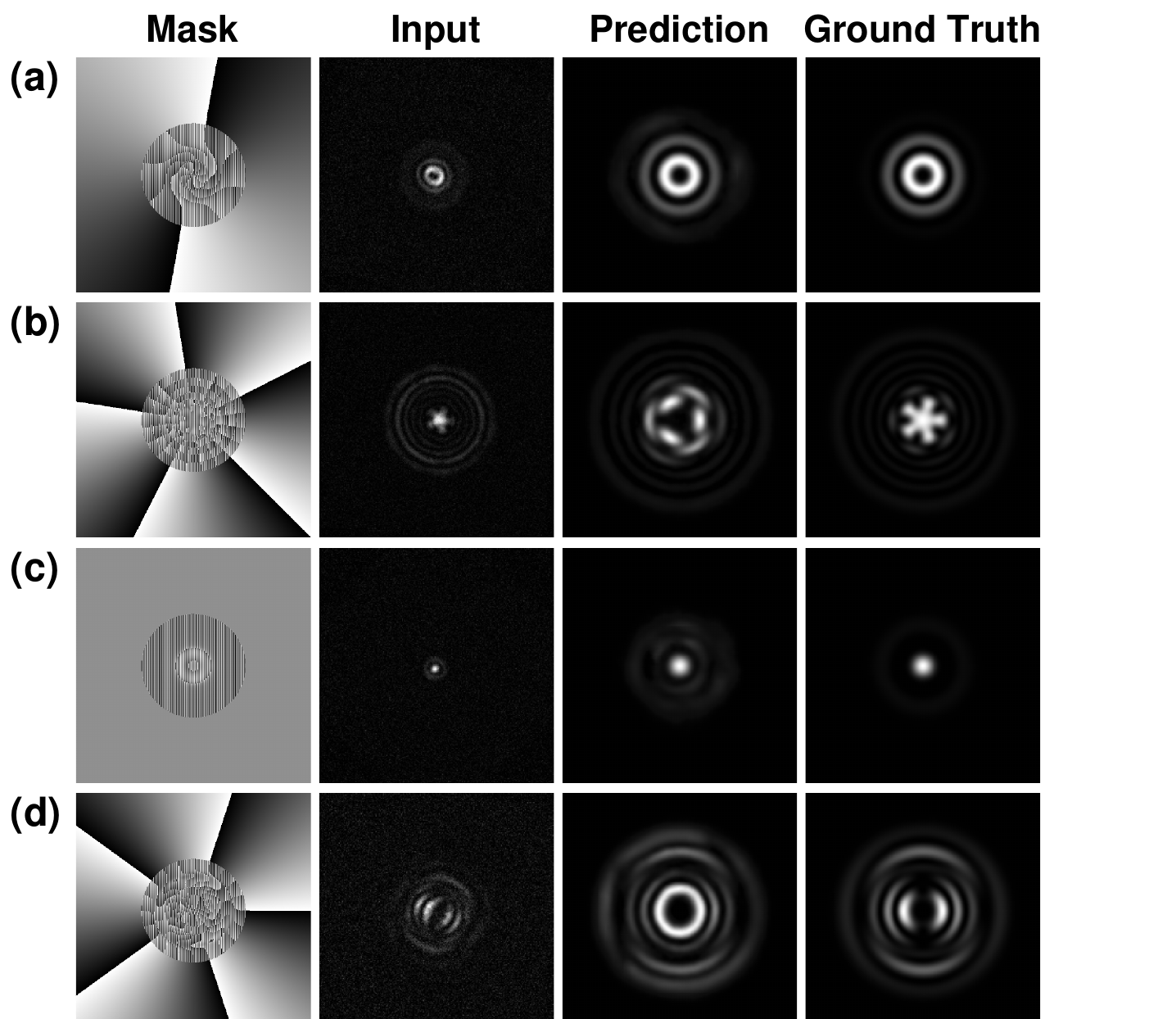}
	\caption{\textbf{Unsuccessful calibration and classification of two-modes superpositions:} Some examples of unsuccessful calibration and classification of the whole BPNet flow. The "Mask" column shows phase masks that where loaded into the SLM.  The "Input" column shows images captured by the camera. The "Prediction" column shows the output of the calibration network while the "Ground Truth" column shows the projected output for a perfect calibration. The superpositions demonstrated and the unsuccessful predictions are : \textbf{(a)} ((1,2),(2,2)) \protect$\to$ ((1,2),(2,2),(3,2)) \textbf{(b)} ((5,5),(0,0)) \protect$\to$ ((5,5),(0,2)) \textbf{(c)} ((0,0),(2,0)) \protect$\to$ ((1,0),(2,1)) \textbf{(d)} ((4,5),(2,3)) \protect$\to$ ((4,5)).
	}
	\label{fig:unsuccessful_SP2}
\end{figure}

	Examples of some unsuccessful results are shown in Fig.\ref{fig:unsuccessful_SP2}. In these cases, it is clear that the images captured (using an automated procedure) by the camera are distorted and clouded. Even though the results were misclassified for these cases, we can appreciate that the calibration network locked on to some features in the input images. Explicitly, we can observe that in all images (except Fig.\ref{fig:unsuccessful_SP2}(a)), even though the ground truth and the captured image look similar, some additional artifacts were introduced to the image and so the input was misclassified. In Fig.\ref{fig:unsuccessful_SP2}(a) the classification network adds an additional mode to the label of the image, due to a deformation in the outer ring. In Fig. \ref{fig:unsuccessful_SP2}(b) the calibration network converts the input image into a completely different superposition of modes. In Fig.\ref{fig:unsuccessful_SP2}(c) a spurious ring appears again but this time instead of adding another mode it simply increases the OAM index for the first mode and the radial index for the second mode. In Fig.\ref{fig:unsuccessful_SP2}(d) the input looks similar to the ground truth image, but the calibration network (probably due to added noise) predicts an image with a closed inner ring, therefore leading to the omission of one of the superimposed modes. Still, it is noteworthy that, in all cases, the predicted images were close to the ground truth images. 
			
\begin{figure}[htbp]
		\centering
		\includegraphics[width=0.75\linewidth,trim={3cm 12cm 10cm 8cm},clip]{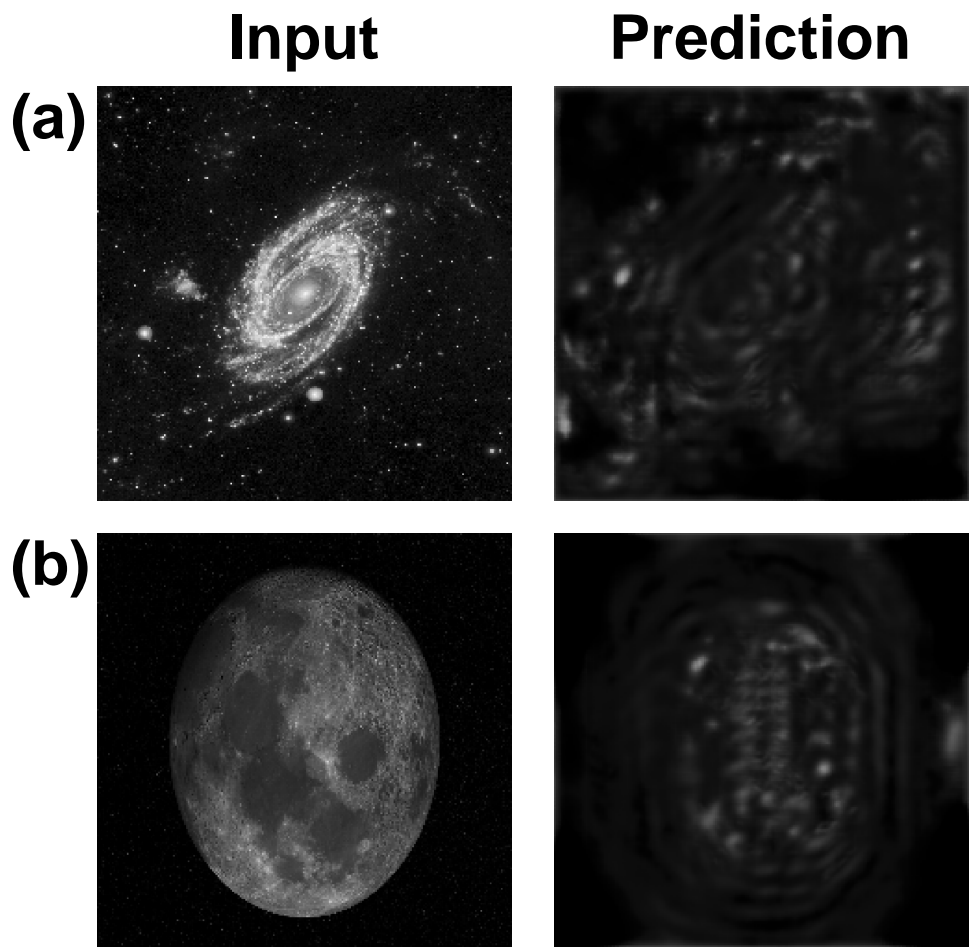}
		\caption{\textbf{Random inputs:} Some examples of random inputs to the calibration network and their predictions.
		}
		\label{fig:random_input}
	\end{figure}

	Finally, it can be suggested that the calibration network did not actually learn to transform its input images to undistorted images (undoing the optical aberrations in the experimental setup)  but that it simply learned a mapping between the input and output images. To refute this argument we fed a few random images to the calibration network (see Fig. \ref{fig:random_input}) where we observe that the output of the  calibration network would not simply map any image with some features to a multiplexed mode, but instead it does learn some kind of a transfer function (albeit it might be restricted to work correctly with the LG modes fed to the network).

	\section{Conclusions}
	We have realized a novel method for spatial-modes de-multiplexing, relating to the two topological numbers characterizing  Laguerre-Gaussian modes using a flow of two concatenated deep neural networks: a calibration network (transferring from experimentally acquired images in the lab to "ideal" images) and a classifier network. We have shown that our classifier is able to demux up to three superimposed spatial-modes with perfect accuracy, while we demonstrated that the whole flow exhibits state-of-the-art performance for detecting two-mode superpositions acquired in the lab. An important ingredient in this work is the introduction of the "Histogram Weighted Loss" which helps handle sparse images where most pixels do not carry information. This loss function might be relevant to other fields that encounter sparse images such as medical imaging and astrophysics.

	\begin{table}[h!]
		\centering
		\begin{tabular}{ | m{7em} | m{6em}| m{6em} | m{3em} | } 
			\hline
			Test-Set & Calibration network DB & Classification network DB & Results \\
			\hline
			Single Mode & $DB_1^{exp}$ & $DB_1^{num}$ & $100\%$ \\
			\hline
			Single Mode & $DB_1^{exp}$ & $DB_2^{num}$ & $96.39\%$ \\
			\hline
			Single Mode & $DB_1^{exp}$ & $DB_3^{num}$ & $86.39\%$ \\
			\hline
			2 Modes Superposition & $DB_2^{exp}$ & $DB_2^{num}$ & $91.3\%$ \\
			\hline
			2 Modes Superposition & $DB_2^{exp}$ & $DB_3^{num}$ & $63.45\%$ \\
			\hline
		\end{tabular}
	\caption{Results for different training configurations. "DB" stands for the training set being used as detailed in the text.}
	\label{table: results}
	\end{table}
	
	\pagebreak
	
	\bibliographystyle{unsrt}
	\bibliography{sample}

\begin{thebibliography}{10}

\bibitem{wang2012terabit}
Jian Wang, Jeng-Yuan Yang, Irfan~M Fazal, Nisar Ahmed, Yan Yan, Hao Huang,
  Yongxiong Ren, Yang Yue, Samuel Dolinar, Moshe Tur, et~al.
\newblock Terabit free-space data transmission employing orbital angular
  momentum multiplexing.
\newblock {\em Nature photonics}, 6(7):488, 2012.

\bibitem{nagali2009quantum}
Eleonora Nagali, Fabio Sciarrino, Francesco De~Martini, Lorenzo Marrucci, Bruno
  Piccirillo, Ebrahim Karimi, and Enrico Santamato.
\newblock Quantum information transfer from spin to orbital angular momentum of
  photons.
\newblock {\em Physical review letters}, 103(1):013601, 2009.

\bibitem{bozinovic2013terabit}
Nenad Bozinovic, Yang Yue, Yongxiong Ren, Moshe Tur, Poul Kristensen, Hao
  Huang, Alan~E Willner, and Siddharth Ramachandran.
\newblock Terabit-scale orbital angular momentum mode division multiplexing in
  fibers.
\newblock {\em science}, 340(6140):1545--1548, 2013.

\bibitem{krenn2014generation}
Mario Krenn, Marcus Huber, Robert Fickler, Radek Lapkiewicz, Sven Ramelow, and
  Anton Zeilinger.
\newblock Generation and confirmation of a (100$\times$ 100)-dimensional
  entangled quantum system.
\newblock {\em Proceedings of the National Academy of Sciences},
  111(17):6243--6247, 2014.

\bibitem{lightman2017miniature}
Shlomi Lightman, Gilad Hurvitz, Raz Gvishi, and Ady Arie.
\newblock Miniature wide-spectrum mode sorter for vortex beams produced by 3d
  laser printing.
\newblock {\em Optica}, 4(6):605--610, 2017.

\bibitem{doster2017machine}
Timothy Doster and Abbie~T Watnik.
\newblock Machine learning approach to oam beam demultiplexing via
  convolutional neural networks.
\newblock {\em Applied optics}, 56(12):3386--3396, 2017.

\bibitem{lohani2018use}
Sanjaya Lohani, Erin~M Knutson, Matthew O’Donnell, Sean~D Huver, and Ryan~T
  Glasser.
\newblock On the use of deep neural networks in optical communications.
\newblock {\em Applied optics}, 57(15):4180--4190, 2018.

\bibitem{gu2018gouy}
Xuemei Gu, Mario Krenn, Manuel Erhard, and Anton Zeilinger.
\newblock Gouy phase radial mode sorter for light: Concepts and experiments.
\newblock {\em Physical review letters}, 120(10):103601, 2018.

\bibitem{zhou2017sorting}
Yiyu Zhou, Mohammad Mirhosseini, Dongzhi Fu, Jiapeng Zhao, Seyed
  Mohammad~Hashemi Rafsanjani, Alan~E Willner, and Robert~W Boyd.
\newblock Sorting photons by radial quantum number.
\newblock {\em Physical review letters}, 119(26):263602, 2017.

\bibitem{bouchard2018measuring}
Fr{\'e}d{\'e}ric Bouchard, Natalia~Herrera Valencia, Florian Brandt, Robert
  Fickler, Marcus Huber, and Mehul Malik.
\newblock Measuring azimuthal and radial modes of photons.
\newblock {\em Optics express}, 26(24):31925--31941, 2018.

\bibitem{chollet2015keras}
Fran\c{c}ois Chollet et~al.
\newblock Keras.
\newblock \url{https://keras.io}, 2015.

\bibitem{ronneberger2015u}
Olaf Ronneberger, Philipp Fischer, and Thomas Brox.
\newblock U-net: Convolutional networks for biomedical image segmentation.
\newblock In {\em International Conference on Medical image computing and
  computer-assisted intervention}, pages 234--241. Springer, 2015.

\bibitem{lin2017focal}
Tsung-Yi Lin, Priya Goyal, Ross Girshick, Kaiming He, and Piotr Doll{\'a}r.
\newblock Focal loss for dense object detection.
\newblock In {\em Proceedings of the IEEE international conference on computer
  vision}, pages 2980--2988, 2017.

\bibitem{sandler2018mobilenetv2}
Mark Sandler, Andrew Howard, Menglong Zhu, Andrey Zhmoginov, and Liang-Chieh
  Chen.
\newblock Mobilenetv2: Inverted residuals and linear bottlenecks.
\newblock In {\em Proceedings of the IEEE Conference on Computer Vision and
  Pattern Recognition}, pages 4510--4520, 2018.

\end{thebibliography}
	
\end{document}